\documentclass[a4paper, UKenglish, cleveref, autoref, thm-restate, numberwithinsect]{lipics-v2021}
\pdfoutput=1

\hideLIPIcs

\title{Learning Aggregate Queries Defined by\texorpdfstring{\\}{} First-Order Logic with Counting\footnote{This is the extended version of the conference contribution \cite{vanBergeremSchweikardt_AggregateQueries}.}}

\titlerunning{Learning Aggregate Queries Defined by First-Order Logic with Counting}

\author{Steffen {van Bergerem}}{Humboldt-Universität zu Berlin, Germany}{steffen.van.bergerem@informatik.hu-berlin.de}{https://orcid.org/0000-0002-5212-8992}{}

\author{Nicole Schweikardt}{Humboldt-Universität zu Berlin, Germany}{schweikn@informatik.hu-berlin.de}{https://orcid.org/0000-0001-5705-1675}{}

\authorrunning{Steffen van Bergerem and Nicole Schweikardt}

\Copyright{Steffen van Bergerem and Nicole Schweikardt}



\category{}

\relatedversion{}

\funding{This work was funded by the Deutsche Forschungsgemeinschaft (DFG, German Research Foundation) --- project number 541000908 (gefördert durch die Deutsche Forschungsgemeinschaft (DFG) --- Projektnummer 541000908).}

\nolinenumbers

\EventEditors{John Q. Open and Joan R. Access}
\EventNoEds{2}
\EventLongTitle{42nd Conference on Very Important Topics (CVIT 2016)}
\EventShortTitle{CVIT 2016}
\EventAcronym{CVIT}
\EventYear{2016}
\EventDate{December 24--27, 2016}
\EventLocation{Little Whinging, United Kingdom}
\EventLogo{}
\SeriesVolume{42}
\ArticleNo{23}

\usepackage[textsize=footnotesize]{todonotes}
\usepackage{framed}
\usepackage{mathtools} \usepackage{stmaryrd}

\newcommand{\ie}{i.\,e.}
\newcommand{\Ie}{I.\,e.}
\newcommand{\eg}{e.\,g.}

\newcommand{\NN}{\mathbb{N}}
\newcommand{\NNpos}{\ensuremath{\NN_{\scriptscriptstyle \geq 1}}}

\newcommand{\ZZ}{\mathbb{Z}}

\newcommand{\Structure}[1]{\ensuremath{\mathcal{#1}}}
\newcommand{\SA}{\Structure{A}}
\newcommand{\SB}{\Structure{B}}

\newcommand{\Class}[1]{\ensuremath{\mathcal{#1}}}

\newcommand{\CG}{\Class{G}}

\newcommand{\I}{\ensuremath{\mathcal{I}}}

\newcommand{\deff}{\coloneqq}

\newcommand{\neighb}[3]{\ensuremath{N_{#1}^{#2}(#3)}} \newcommand{\neighbr}[2]{\neighb{r}{#1}{#2}} \newcommand{\Neighb}[3]{\ensuremath{\mathcal{N}_{#1}^{#2}(#3)}} \newcommand{\Neighbr}[2]{\Neighb{r}{#1}{#2}} \newcommand{\Neighbrt}[2]{\Neighb{r_t}{#1}{#2}} \newcommand{\NeighbrStrich}[2]{\Neighb{r'}{#1}{#2}} 

  \newcommand{\tupEqInSet}[1]{\Rrightarrow_{#1}}

\newcommand{\abs}[1]{\left\lvert#1\right\rvert}
\newcommand{\bigabs}[1]{\bigl\lvert#1\bigr\rvert}

\DeclareMathOperator*{\ar}{ar}

\DeclareMathOperator{\dist}{dist}
\DeclareMathOperator*{\free}{free}

\renewcommand{\leq}{\leqslant}
\renewcommand{\geq}{\geqslant}

\renewcommand{\phi}{\varphi}
\renewcommand{\epsilon}{\varepsilon}

\newcommand{\set}[1]{\ensuremath{\{#1\}}}
\newcommand{\setc}[2]{\ensuremath{\set{#1 : #2}}}
\newcommand{\bigset}[1]{\ensuremath{\bigl\{ #1 \bigr\}}}
\newcommand{\bigsetc}[2]{\bigset{#1 \bigmid #2}}

\newcommand{\bigmid}{\ \mathrel{:} \ }

\newcommand{\emptytuple}{\ensuremath{()}}

\newcommand{\vars}{\ensuremath{\textsf{\upshape vars}}}

\newcommand{\sem}[1]{\left\llbracket #1 \right\rrbracket} \newcommand{\Land}{\ensuremath{\bigwedge}}

\newcommand{\FOCCount}[2]{\ensuremath{\# {#1}.{#2}}}

\newcommand{\bigO}{\mathcal{O}}

\newcommand{\ComplexityClassFont}[1]{\ensuremath{\textup{\textsf{#1}}}}

\newcommand{\AWstar}{\ensuremath{\ComplexityClassFont{AW}[*]}}

\newcommand{\LogicFont}[1]{\ComplexityClassFont{#1}}
\newcommand{\FO}{\LogicFont{FO}}
\newcommand{\MSO}{\LogicFont{MSO}}
\newcommand{\FOCN}{\LogicFont{FOCN}}

\newcommand{\FOC}{\LogicFont{FOC}}

\newcommand{\FOunC}{\ensuremath{\FOC_1}}

\newcommand{\FOWA}{\LogicFont{FOWA}} \newcommand{\FOWAun}{\ensuremath{\FOWA_1}}

\newcommand{\PP}{\ensuremath{\mathbb{P}}}
\newcommand{\Pred}{\ensuremath{\mathsf{P}}}

\renewcommand{\phi}{\varphi}

\newcommand{\LearnWithPrecomp}{\ensuremath{\textup{\textsc{Learn-with-Precomputation}}(k, \ell, T, T')}}

\newcommand{\ov}[1]{\ensuremath{\bar{#1}}}
\newcommand{\A}{\SA}

\begin{document}

\maketitle

\begin{abstract}
  In the logical framework introduced by Grohe and Tur{\'{a}}n (TOCS~2004) for
Boolean classification problems, the instances to classify are tuples from
a logical structure, and Boolean classifiers are described by parametric models
based on logical formulas. This is a specific scenario for supervised passive learning,
where classifiers should be learned based on labelled examples.
Existing results in this scenario focus on Boolean classification.
This paper presents learnability results beyond Boolean
classification.
We focus on multiclass classification
problems where the task is to assign input tuples to arbitrary integers.
To represent such integer-valued classifiers, we
use aggregate queries specified by an extension of first-order
logic with counting terms called $\FOunC$.

Our main result shows the following:
given a database of polylogarithmic degree,
within quasi-linear time,
we can build an index structure that makes it possible to learn
\(\FOunC\)-definable integer-valued classifiers
in time polylogarithmic in the size of the database
and polynomial in the number of training examples.
 \end{abstract}

\section{Introduction}
\label{sec:intro}

We study the complexity of learning aggregate queries from examples.
This is a \emph{classification problem} of the following form.
The elements that are to be classified come from a set \(X\), the \emph{instance space}.
For a given set \(V\), a \(V\)-valued \emph{classifier} on \(X\) is a function \(c \colon X \to V\).
We are given a \emph{training set} \(S\) of labelled examples \((x,\lambda) \in X\times V\),
\ie, \(\lambda\) is the label assigned to the instance \(x\).
The goal is to find a classifier, called a \emph{hypothesis},
that can be used to predict the label of elements from \(X\), including those not given in \(S\).
The term \emph{Boolean classification problem} refers to the case where \(\abs{V} = 2\)
(often, \(V\) is \(\set{1, 0}\)).
We use the term \emph{multiclass classification problem} to refer to cases where \(V\) may be arbitrarily large.
In machine learning, these
problems fall into the category of \emph{supervised learning} tasks:
we want to learn a function from given input-output pairs.
In contrast to this, in unsupervised learning (\eg\ clustering),
the goal is to learn patterns from unlabelled data
\cite{Shalev-ShwartzBen-David_UnderstandingMachineLearning}.

We focus on learning problems related to the framework
introduced by Grohe and Tur{\'{a}}n~\cite{GroheTuran_Learnability}.
There, the instance space \(X\) is a set of tuples from a logical
structure (that is sometimes called the \emph{background structure}),
and the classifiers are Boolean and are described using parametric models based on
logical formulas.
In this paper, we extend the framework to multiclass classification problems
where the classifiers are integer-valued, \ie, \(V = \ZZ\).
In the framework that we consider,
the background structure is a relational database \(\SA\),
and the instance space \(X\) is the set \(A^k\) of all \(k\)-tuples of elements
from the active domain \(A\) of \(\SA\) (also called the \emph{universe} of \(\SA\)).
Here, $k$ is a fixed positive integer.
One fixes a \emph{parameter length} \(\ell\) (a fixed non-negative integer).
A classifier is specified by a pair \(p=(t,\ov{w})\),
where $\ov{w}=(w_1, \dots, w_\ell)$ is an $\ell$-tuple of elements in $A$,
and \(t\) is a \emph{counting term} in the \emph{first-order logic with counting}
\(\FOunC\) \cite{GroheSchweikardt_FOunC}
with free variables \(x_1,\dots, x_k, y_1, \dots, y_\ell\).
This pair $p$ represents the classifier $c_p\colon X\to \ZZ$
that assigns to each $k$-tuple \(\ov{a} = (a_1, \dots, a_k) \in X\) the integer \(i\)
that is obtained by evaluating the counting term \(t\)
in the database \(\SA\) while interpreting the variables
\(x_1, \dots, x_k\) with the elements \(a_1, \dots, a_k\)
and the variables \(y_1, \dots, y_\ell\) with the ``parameters'' \(w_1, \dots, w_\ell\).
We will write \(h^\SA_{t,\ov{w}}\) to denote this classifier $c_p$.

Given a \emph{training set} \(S \subseteq A^k \times \ZZ\),
we want to find a pair $p=(t,\ov{w})$
such that the classifier \(h^\SA_{t,\bar{w}}\) is \emph{consistent} with \(S\),
\ie, it satisfies \(h^\SA_{t, \bar{w}}(\bar{a}) = i\) for every \((\bar{a}, i) \in S\).

\begin{example}
  \label{ex:citations}
  Let \(\SA\) be a relational database
  where the active domain \(A\) contains authors and publications,
  the binary relation \texttt{Author} contains all pairs \((a,p)\)
  where \(a\) is an author of the publication \(p\),
  and the binary relation \texttt{Citation} contains all pairs \((p_1, p_2)\)
  where the publication \(p_1\) cites the publication \(p_2\).

  Suppose we are given a training set \(S\)
  that consists of a few pairs \((a, i)\)
  where \(a\) is an author and \(i\) is the total number of citations
  of the publications of \(a\).
A reasonable classifier for this setting would be a mapping
  $c \colon  A \to \ZZ$ that assigns to every author $a$ present in the
  database the total number $i$ of citations of their publications.
In our setting, this can be represented as follows.
  We let \(k = 1\) and \(\ell = 0\).
  Since $\ell=0$, the ``parameter'' $w$ is fixed to be the empty tuple $\emptytuple$.
  Since $k=1$, we use a counting term with a single free variable $x$
  (that will be assigned with authors present in the database). We choose the
  counting term
  \[
    t(x) \ \deff \ \ \FOCCount{(z_1, z_2)}{\bigl(\texttt{Author}(x,z_1)
      \land \texttt{Citation}(z_2, z_1)\bigr)}.
  \]
  Evaluating $t(x)$ for an author $x$ yields the number of
  tuples $(z_1,z_2)$ such that $x$ is an author of publication $z_1$,
  and $z_2$ is a publication that cites $z_1$. This is precisely the
  total number of citations of publications authored by $x$.
  Hence, $h^{\A}_{t,\emptytuple}$ is the desired classifier $c$.
\end{example}

\begin{example}
  \label{ex:cake}
  Suppose we have a database that maintains a list of all cakes colleagues brought to work.
  We model this as a relational database \(\SA\)
  whose active domain \(A\) contains persons, IDs of cakes, and types of cake.
  The binary relation $\texttt{Brought}$ contains all pairs \((p,c)\)
  where \(p\) is a person that brought the cake with ID \(c\),
  and the binary relation $\texttt{Type}$ contains all pairs \((c,\tau)\)
  where \(c\) is the ID of a cake of type \(\tau\)
  (\eg, ``chocolate cake'', ``strawberry cake'', ``carrot cake'', etc).
Suppose we want to find a classifier that predicts the popularity of colleagues.
  For this, via a survey, we gather examples \((p, i) \in A \times \ZZ\)
  where \(p\) is a person and \(i\) is the popularity of the person,
  and we call the resulting set of labelled examples \(S\).
  We choose \(k = \ell = 1\),
  so we want to find a classifier that uses a single parameter.
  According to our own experience at work, it seems conceivable that
  the following classifier \(h^\SA_{t,w}\) is consistent with \(S\):
  the parameter $w$  is “chocolate cake” and
  \(t\) is the counting term
  \[t(x,y) \ \deff \ \ \FOCCount{(z)}{\bigl(
    \texttt{Brought}(x,z) \land \neg \texttt{Type}(z,y)\bigr)}
    \ \ + \ \ 2 \cdot \FOCCount{(z)}{\bigl(\texttt{Brought}(x,z) \land
    \texttt{Type}(z,y)\bigr)}.
  \]
  Note that $t$ counts the number of cakes brought by person \(x\),
  where cakes of type \(y\) are counted twice,
  and the variable $y$ will always be assigned the value of the parameter $w$.
\end{example}

In many application scenarios, the same database is used multiple times
with different training sets to learn different classifiers.
Thus, we consider a setting in which we are first only given the database,
without any training examples.
In a precomputation step, we allow gathering information
that will be helpful for solving future learning tasks.
This precomputation step can be viewed as building an index structure
that is designed in order to support solving multiple learning tasks.

In the actual learning phase, we are repeatedly given training sets of labelled examples,
and our task is to output a hypothesis that is consistent with the corresponding training set.
For this learning phase, it would be desirable to have algorithms that run efficiently
even if the database is too large to fit into the main memory.
To achieve this, we are interested in algorithms that require only \emph{local access} to the database,
\ie, instead of having random access to the database,
a learning algorithm should initially start with the elements given in the training set;
subsequently, it may only retrieve the neighbours of elements it already holds in memory.
By utilising the memory hierarchy, such local access can be achieved
efficiently even in cases where random access is too prohibitive.
In the context of learning (concerning Boolean classification problems),
this local-access model has been introduced by Grohe and Ritzert~\cite{GroheRitzert_FO}.

\paragraph*{Our contribution}
Our main result is an algorithm that builds the index structure in time
linear in the size and polynomial in the degree of the database.
Afterwards, upon input of concrete training sets,
classifiers definable in \(\FOunC\)
can be learned in time polynomial in the degree of the database
and polynomial in the number of examples given in the training set.
Moreover, the classifiers returned by our algorithm
can be evaluated in time polynomial in the degree of the database.
Furthermore, our algorithms for finding a classifier and for
evaluating this classifier do not require random access to the
database but only rely on the local-access model.

For databases of polylogarithmic degree
(\ie, of degree up to $(\log n)^c$ where $c$ is a constant and $n$ is the size of the database),
our main result implies that the index structure can be built in quasi-linear time
(\ie, time $n{\cdot}(\log n)^c$);
afterwards, \(\FOunC\)-definable integer-valued classifiers can be learned in time
polylogarithmic (so, in particular, sublinear) in the size of the database
and polynomial in the number of training examples.

Previous results in the framework of Grohe and Tur{\'{a}}n
for Boolean classification problems relied on the fact
that it suffices to check a constant number of queries
while limiting the search space for the parameters to a neighbourhood of a certain radius
\cite{GroheRitzert_FO,vanBergerem_FOCN,vanBergeremSchweikardt_FOWA}.
For our setting of multiclass classification with aggregate queries, however,
this does not hold any more.
Hence, a priori, it is not clear that sublinear-time learning algorithms
are possible for the multiclass case at all.
The main technical challenge towards our learnability result
was to find an approach that keeps the number of queries to check small
(\ie, polynomial in the degree of the database),
while still being able to limit the search space for the parameters
to a small neighbourhood around the given training tuples.

\paragraph*{Organisation}
We provide the necessary definitions concerning $\FOunC$ in
Section~\ref{sec:prelim}, and
we formally introduce the learning problem that we consider in \cref{sec:learning}.
The precise statement of our main result is given in \cref{thm:learn-foc1}.
Our proof makes heavy use of the locality properties of the logic \(\FOunC\)
shown in \cite{GroheSchweikardt_FOunC},
including a decomposition of \(\FOunC\)-formulas into local formulas.
These properties are used in \cref{sec:proof-of-main-lemma}
to provide our main technical tool for the proof of the main result.
Section~\ref{sec:conclusion} concludes the paper with a summary and an
outlook on future work.
In the remainder of this introduction,
we give an overview of related work.

\paragraph*{Related work}
The first-order logic with counting $\FOC$
was introduced in  \cite{KuskeSchweikardt_FOCN}
and further studied in
\cite{GroheSchweikardt_FOunC,vanBergerem_FOCN}.
This logic extends first-order logic ($\FO$) by the ability to formulate \emph{counting terms}
that evaluate to integers, and by \emph{numerical predicates}
that allow to compare results of counting terms.
It was shown in \cite{KuskeSchweikardt_FOCN}
that the model-checking problem for $\FOC$ is fixed-parameter
tractable on classes of structures of bounded degree.
From \cite{GroheSchweikardt_FOunC}
it is known that the fixed-parameter tractability of $\FOC$ cannot be
generalised to even very simple classes of structures of unbounded degree such as
unranked trees (under a reasonable assumption in parameterised complexity).
However, \cite{GroheSchweikardt_FOunC} identified a fragment called $\FOunC$
for which model-checking of formulas and evaluation of counting
terms are fixed-parameter tractable on all nowhere dense classes of
structures.
The present paper uses counting terms of $\FOunC$ to represent
integer-valued classifiers.

The learning framework we consider has been introduced
for Boolean classification problems in \cite{GroheTuran_Learnability},
which provides information-theoretic learnability results
for classes of classifiers that can be specified using \(\FO\)- and \(\MSO\)-formulas
on restricted classes of structures,
such as the class of planar graphs or classes of graphs of bounded degree.
Algorithmic aspects of the framework,
including the running time of a learning algorithm,
were first studied
in \cite{GroheRitzert_FO}.
The paper showed that Boolean classifiers definable in \(\FO\)
can be learned in sublinear time on structures of polylogarithmic degree.
Analogous results have been obtained for \(\MSO\)
on strings \cite{GroheLoedingRitzert_MSO}
and on trees \cite{GrienenbergerRitzert_Trees},
which included a precomputation step to allow for efficient repeated learning.
The paper \cite{vanBergeremGroheRitzert_Parameterized}
studied the parameterised complexity
of the Boolean classification problem and
showed that on arbitrary relational structures,
learning hypotheses definable in $\FO$ is hard for
the parameterised complexity class \(\AWstar\) (\ie, subject to a
plausible complexity-theoretic assumption, it is not fixed-parameter tractable).
The paper also showed that the problem is fixed-parameter tractable
if the structures come from a nowhere dense class.

For Boolean classifiers definable in the extension \(\FOCN\) of \(\FO\) with
counting quantifiers and numerical predicates,
\cite{vanBergerem_FOCN} obtained a sublinear-time learning algorithm
for {structures of bounded degree}, \ie,
classes of structures where the degree is bounded by a constant.
Recently, \cite{vanBergerem_PhDThesis}
lifted this result to structures of tiny degree, \ie,
classes of structures of degree up to
\((\log \log n)^c\) for some constant \(c\),
where \(n\) is the size of the structure.
The paper \cite{vanBergeremSchweikardt_FOWA} considered a notion of
\emph{weighted structures}, which extend
ordinary relational structures by assigning weights, \ie\ elements
from particular rings or abelian groups, to tuples present in the structure.
It introduced the expressive logic $\FOWA$, which extends \(\FO\)
by means of aggregating weights and formulating both
formulas (that evaluate to ``true'' or ``false'') and
terms (that ``aggregate'' weights and evaluate to values in the
associated ring or abelian group). For the fragment $\FOWAun$ (that
still extends $\FO$), the paper showed that
Boolean classifiers definable by $\FOWAun$-formulas over weighted
background structures of polylogarithmic degree can be learned
in sublinear time after quasi-linear-time preprocessing.
This lifts the results obtained in \cite{GroheRitzert_FO} for $\FO$
to the substantially more expressive logic $\FOWAun$.
As the logic $\FOunC$ can be embedded in $\FOWAun$, it follows from
\cite{vanBergeremSchweikardt_FOWA} that
Boolean classifiers definable by $\FOunC$-formulas
over background structures of polylogarithmic degree
can be learned in sublinear time after quasi-linear-time preprocessing.
The main result of the present paper can be viewed as a generalisation of
this to integer-valued classification problems.

The algorithmic results obtained so far
within the framework introduced in \cite{GroheTuran_Learnability}
all focus on \emph{Boolean classification} problems.
However, many application scenarios require \emph{multiclass classification}
(cf.\ \cite{DBLP:conf/nips/DanielySS12,DBLP:conf/focs/BrukhimCDMY22,DBLP:conf/colt/HannekeMZ23}).
In the database systems literature, multiclass classifiers typically are
described by \emph{aggregate queries}
\cite{TanZhangElmeleegySrivastava_REGAL2017,TanZhangElmeleegySrivastava_REGAL2018,TranChanParthasarathy_ReverseEngineering2014,WangCheungBodik_SynthesizingSQL2017,Martins_ReverseEngineering2019}.
In this paper, aggregate queries are represented by the counting terms of $\FOunC$.

Closely related to the framework we consider is the framework of
\emph{inductive logic programming} (ILP)
\cite{Muggleton_ILP91,MuggletonDeRaedt_ILP94,KietzDzeroski_ILP,CohenPage_ILP95,CropperDumancicEvansMuggleton_ILP2022}.
Both frameworks deal with a \emph{passive supervised learning} setting,
where the learning algorithms are given labelled examples.
These examples are labelled according to some target concept, and the algorithms
should return a hypothesis that approximately matches this target concept.
One of the main differences between both frameworks is that we represent
the background knowledge by a relational database,
whereas in ILP, it is represented in a background theory,
\ie, a set of formulas.
Related logical learning frameworks have also been studied in formal verification
\cite{GargLoedingMadhusudanNeider_Verification2014,LoedingMadhusudanNeider_Verification2016,EzudheenNeiderDSouzaMadhusudan_Verification2018,ZhuMagillJagannathan_Verification2018,ChampionChibaKobayashiSato_Verification2020}.

In the database literature, various approaches to learning queries from examples
have been studied, both in passive (such as ours) and active learning settings.
In passive learning settings,
results often focus on conjunctive queries~\cite{Haussler_ConjunctiveConcepts1989,Hirata_AcyclicConjunctiveQueries2000,BarceloRomero_ConjunctiveQueries2017,KimelfeldRe_RelationalFramework2018,BarceloBDKimelfeld_ConjunctiveQueries2021,WeissCohen_SPJQueries2017}
or consider queries outside the relational database model~\cite{StaworkoWieczorek_TwigAndPathQueries2012,BonifatiCiucanuLemay_PathQueries2015},
while we focus on $\FOunC$, an extension of full first-order logic.
In the \emph{active learning} setting
introduced by Angluin~\cite{Angluin_ExactLearning},
learning algorithms are allowed to actively query an oracle.
Results in this setting~\cite{AizensteinHegeduesHellersteinPitt_QueryLearning1998,SloanSzoerenyiTuran_LearningBooleanFunctionsWithQueries2010,AbouziedAngluinPHS_LearningBooleanQueries2013,BonifatiCiucanuLemay_PathQueries2015,
BonifatiCiucanuStaworko_JoinQueries2016,
tenCateDalmau_ConjunctiveQueries2021}
again consider various types of queries.
Another related subject in the database literature
is the problem of learning schema mappings from examples
\cite{BonifatiComignaniCoqueryThion_SchemaMapping2019,GottlobSenellart_SchemaMappings2010,AlexeTenCateKolaitisTan_SchemaMappings2011,tenCateDalmauKolaitis_SchemaMappings2013,tenCateKolaitis_SchemaMappings2018}.

\section{Preliminaries}
\label{sec:prelim}
This section fixes the basic notation used throughout the paper, and
it provides the precise syntax and semantics of first-order logic with
counting $\FOunC$ (the latter is taken almost verbatim from \cite{GroheSchweikardt_FOunC}).
We let \(\ZZ\), \(\NN\), and \(\NNpos\) denote the sets of
integers, non-negative integers, and positive integers, respectively.
For \(m, n \in \ZZ\),
we let \([m,n] \deff \setc{\ell \in \ZZ}{m \leq \ell \leq n}\)
and \([n] \deff [1,n]\).
For a $k$-tuple $\ov{x}=(x_1,\ldots,x_k)$ we write $|\ov{x}|$ to
denote its \emph{arity} $k$. By $\emptytuple$ we denote the empty
tuple, \ie, the tuple of arity $0$.

\paragraph*{Signatures and structures}
A \emph{signature} is a finite set of relation symbols.
Every relation symbol \(R\) has a fixed \emph{arity} \(\ar(R) \in \NN\).
Let \(\sigma\) be a signature.
A \emph{\(\sigma\)-structure} (or \emph{$\sigma$-database}) \(\SA\)
consists of a finite set $A$, called the \emph{universe} of $\A$,
and a relation \(R(\SA) \subseteq A^{\ar(R)}\)
for every \(R \in \sigma\).
Note that signatures may contain
relation symbols of arity 0. There are only two 0-ary relations over a
set $A$, namely $\emptyset$ and $\set{\emptytuple}$, which we
interpret as ``false'' and ``true'', respectively.
We define the \emph{size} $\abs{\SA}$ of a $\sigma$-structure $\A$ as
\(\abs{\SA} \deff |A|\).

Let $\sigma'$ be a signature with \(\sigma' \supseteq \sigma\).
A \emph{\(\sigma'\)-expansion} of a $\sigma$-structure $\A$ is
a \(\sigma'\)-structure \(\SA'\)
which has the same universe as $\A$
and which satisfies \(R(\SA') = R(\SA)\) for all \(R \in \sigma\).

A \emph{substructure} of a $\sigma$-structure $\A$ is
a \(\sigma\)-structure \(\SB\) with universe $B\subseteq A$
that satisfies
\(R(\SB) \subseteq R(\SA)\) for every \(R \in \sigma\).
For a set \(X \subseteq A\), the \emph{induced substructure of \(\SA\) on \(X\)}
is the \(\sigma\)-structure \(\SA[X]\) with universe \(X\), where
\(R(\SA[X]) = R(\SA) \cap X^{\ar(R)}\) for every \(R \in \sigma\).

\paragraph*{Gaifman graph, degree, distances, and neighbourhoods}
In this paper, when speaking of graphs, we mean undirected simple graphs.

The \emph{Gaifman graph} \(G_\SA\) of a \(\sigma\)-structure \(\SA\) with universe \(A\)
is the graph with vertex set \(V(G_{\SA})=A\) whose edge set \(E(G_\SA)\)
contains exactly those edges $\set{v,w}$ where $v,w \in A$,
$v\neq w$, and there exists an $R\in\sigma$ and a tuple
$(a_1,\ldots,a_{\ar(R)})\in R(\A)$ such that $v,w\in\set{a_1,\ldots,a_{\ar(R)}}$.
The \emph{degree} of a $\sigma$-structure $\A$ is defined as the
degree of the Gaifman graph $G_\A$ (\ie, the maximum number of
neighbours of a vertex of $G_\A$).

The \emph{distance} \(\dist^\SA(a,b)\) between two elements \(a, b \in A\)
is the minimal number of edges of a path from \(a\) to \(b\) in \(G_\SA\);
if no such path exists, we let \(\dist^\SA(a,b) \deff \infty\).
For a tuple \(\bar{a} = (a_1, \dots, a_k) \in A^k\)
and an element \(b \in A\),
we let \(\dist^\SA(\bar{a}, b) \deff \min_{i \in [k]} \dist^\SA(a_i,
b)\).

Consider a $k$-tuple $\ov{a}=(a_1,\ldots,a_k)\in A^k$ for some $k\in\NNpos$.
For every \(r \geq 0\),
the \emph{ball of radius \(r\)} (or \emph{\(r\)-ball}) of \(\bar{a}\) in \(\SA\)
is the set \(\neighbr{\SA}{\bar{a}} \deff \setc{b \in A}{\dist^\SA(\bar{a}, b) \leq r}\).
The \emph{neighbourhood of radius \(r\)} (or \emph{\(r\)-neighbourhood})
of \(\bar{a}\) in \(\SA\) is the induced substructure
\(\Neighbr{\SA}{\bar{a}} \deff \SA[\neighbr{\SA}{\bar{a}}]\).

\paragraph*{First-order logic with counting $\FOunC$}
Let \(\vars\) be a fixed, countably infinite set
of \emph{variables}.
A \emph{\(\sigma\)-interpretation} \(\I = (\SA, \beta)\) consists of a
\(\sigma\)-structure \(\SA\) and an \emph{assignment}
\(\beta \colon \vars \to A\), where $A$ denotes the universe of $\A$.
For \(k \in \NN\) and
\(k\) distinct variables \(x_1, \dots, x_k \in \vars\)
and elements \(a_1, \dots, a_k \in A\),
we write
\(\I \frac{a_1, \dots, a_k}{x_1, \dots, x_k}\)
for the interpretation
\((\SA, \beta \frac{a_1, \dots, a_k}{x_1, \dots, x_k})\),
where
\(\beta \frac{a_1, \dots, a_k}{x_1, \dots, x_k}\)
is the assignment \(\beta'\) with
\(\beta'(x_i) = a_i\) for every \(i \in [k]\)
and \(\beta'(z) = \beta(z)\) for all
\(z \in \vars \setminus \set{x_1, \dots, x_k}\).

Next, we provide the syntax and semantics of the logic $\FOunC$ \cite{GroheSchweikardt_FOunC}.
This logic allows formulating numerical statements based on
counting terms and numerical predicates.

For the remainder of this paper,
fix a triple \((\PP, \ar, \sem{.})\),
called a \emph{numerical predicate collection},
where \(\PP\) is a finite set of \emph{predicate names},
\(\ar\) assigns an \emph{arity} \(\ar(\Pred) \in \NNpos\)
to each \(\Pred \in \PP\),
and \(\sem{.}\) assigns a \emph{semantics}
\(\sem{\Pred} \subseteq \ZZ^{\ar(\Pred)}\) to each $\Pred\in\PP$.
Examples of numerical predicates are
the \emph{equality predicate}
$\Pred_{=}$ with $\sem{\Pred_{=}}=\setc{(i,i)}{i\in \ZZ}$,
the \emph{comparison predicate}
$\Pred_{\leq}$ with $\sem{\Pred_{\leq}}=\setc{(i,j)}{i,j\in\ZZ,\ i\leq
  j}$, or the \emph{prime number predicate}
$\Pred_{\textsf{prime}}$ with
$\sem{\Pred_{\textsf{prime}}}=\setc{i\in\NN}{i \text{ is a prime
    number}}$.
When analysing the running time of algorithms, we will assume that machines have
access to oracles for evaluating the numerical predicates.
That is, when given
a $\Pred\in\PP$ and
a tuple $(i_1,\ldots,i_{\ar(\Pred)})$ of integers, the oracle takes
time $O(1)$ to answer if $(i_1,\ldots,i_{\ar(\Pred})\in\sem{\Pred}$.

Let $\sigma$ be a signature.
The set of \emph{formulas} and \emph{counting terms} (for short: terms)
of \(\FOunC[\sigma]\) is built according to the following rules.

\begin{enumerate}[(1)]
  \item\label{def:fo-atomic}
    \(x_1{=}x_2\) and \(R(x_1, \dots, x_{\ar(R)})\) are \emph{formulas},\footnote{in particular, if $\ar(R)=0$ then $R()$ is a formula}
    where $R\in\sigma$, and \(x_1, x_2, \dots, x_{\ar(R)}\) are variables.
    We let \(\free(x_1{=}x_2) \deff \set{x_1, x_2}\) and
    \(\free\bigl(R(x_1, \dots, x_{\ar(R)})\bigr) \deff \set{x_1, \dots, x_{\ar(R)}}\).
  \item\label{def:fo-bool}
    If \(\phi\) and \(\psi\) are formulas,
    then \(\neg \phi\) and \((\phi \lor \psi)\) are also \emph{formulas}.
    We let \(\free(\neg \phi) \deff \free(\phi)\) and
    \(\free((\phi \lor \psi)) \deff \free(\phi) \cup \free(\psi)\).
  \item\label{def:fo-exists}
    If \(\phi\) is a formula and \(x \in \vars\),
    then \(\exists x\, \phi\) is a \emph{formula}.
    We let \(\free(\exists x\, \phi) \deff \free(\phi) \setminus \set{x}\).
  \item\label{def:foc-countingterm}
    If \(\phi\) is a formula and
    \(\bar{x} = (x_1, \dots, x_k)\) is a tuple of \(k\) pairwise
    distinct variables, for $k\geq 0$,
    then \(\FOCCount{\bar{x}}{\phi}\) is a \emph{counting term}.
    We let
    \(\free\bigl(\FOCCount{(x_1, \dots, x_k)}{\phi}\bigr) \deff \free(\phi) \setminus \set{x_1, \dots, x_k}\).
  \item\label{def:foc-constterm}
    Every integer \(i \in \ZZ\) is a \emph{counting term}.
    We let \(\free(i) = \emptyset\).
  \item\label{def:foc-plustimesterm}
    If \(t_1\) and \(t_2\) are counting terms,
    then \((t_1 + t_2)\) and \((t_1 \cdot t_2)\) are also \emph{counting
    terms}.

    We let \(\free\bigl((t_1 + t_2)\bigr) \deff \free\bigl((t_1 \cdot
    t_2)\bigr) \deff \free(t_1) \cup \free(t_2)\).
  \item\label{def:foc-P}
    If \(\Pred \in \PP\), \(m = \ar(\Pred)\),
    and \(t_1, \dots, t_m\) are counting terms
    with \(\bigabs{\bigcup_{i=1}^m \free(t_i)} \leq 1\),
    then \(\Pred(t_1, \dots, t_m)\) is a \emph{formula}.
    We let \(\free\bigl(\Pred(t_1, \dots, t_m)\bigr) \deff \bigcup_{i=1}^m \free(t_i)\).
\end{enumerate}

\emph{First-order logic} $\FO[\sigma]$ is the fragment of
$\FOunC[\sigma]$ built by using only the rules
\eqref{def:fo-atomic}--\eqref{def:fo-exists}.
We write \((\phi \land \psi)\) and \(\forall x\, \phi\) as shorthands for
\(\neg(\neg \phi \lor \neg \psi)\) and \(\neg \exists x\, \neg \phi\).
For counting terms \(t_1\) and \(t_2\),
we write \((t_1 - t_2)\) as a shorthand for \(\bigl(t_1 + ((-1) \cdot t_2)\bigr)\).

By \(\FOunC\), we denote the union of all \(\FOunC[\sigma]\)
for arbitrary signatures \(\sigma\).
This applies analogously to \(\FO\).
For $\FOunC$ the semantics for the rules \eqref{def:fo-atomic}--\eqref{def:fo-exists}
are defined in the same way as for $\FO$; the semantics of the
remaining rules are as follows.
Let \(\I = (\SA, \beta)\) be a \(\sigma\)-interpretation, and let $A$
denote the universe of $\SA$.

\begin{enumerate}[(1)]
  \setcounter{enumi}{3}
  \item
    \(\sem{\FOCCount{\bar{x}}{\phi}}^\I =
    \Big|
      \bigsetc{(a_1, \dots, a_k) \in A^k}
         {\sem{\phi}^{\I \frac{a_1, \dots, a_k}{x_1, \dots, x_k}} =
           1}
    \Big|\),
    where \(\bar{x} = (x_1, \dots, x_k)\).
  \item
    \(\sem{i}^\I = i\), for \(i \in \ZZ\).
  \item
    \(\sem{(t_1 + t_2)}^\I = \sem{t_1}^\I + \sem{t_2}^\I\)
    and \(\sem{(t_1 \cdot t_2)}^\I = \sem{t_1}^\I \cdot \sem{t_2}^\I\).
  \item
    \(\sem{\Pred(t_1, \dots, t_m)}^\I = 1\)
    if \((\sem{t_1}^\I, \dots, \sem{t_m}^I) \in \sem{\Pred}\),
    and \(\sem{\Pred(t_1, \dots, t_m)}^\I = 0\) otherwise.
\end{enumerate}

Examples of counting terms in $\FOunC$ and their
intuitive meaning can be found in \cref{ex:citations,ex:cake}.

An \emph{expression} is a formula or a counting term.
For an expression $\xi$,
we write \(\xi(z_1, \dots, z_k)\) to indicate that
\(\free(\xi) \subseteq \set{z_1, \dots, z_k}\).
A \emph{sentence} is a formula without free variables.
A \emph{ground term} is a counting term without free variables.

For a formula \(\phi\) and a \(\sigma\)-interpretation \(\I\),
we write \(\I \models \phi\) to indicate that \(\sem{\phi}^\I = 1\).
Likewise, \(\I \not\models \phi\) indicates that
\(\sem{\phi}^\I = 0\).
For a formula \(\phi(x_1, \dots, x_k)\), a \(\sigma\)-structure \(\SA\),
and a tuple \(\bar{a} = (a_1, \dots, a_k) \in A^k\),
we write \(\SA \models \phi[\bar{a}]\)
to indicate that \((\SA, \beta) \models \phi\) for all assignments \(\beta\)
with \(\beta(x_i) = a_i\) for all \(i \in [k]\).
Similarly, for a counting term \(t(x_1, \dots, x_k)\),
we write \(t^\SA[\bar{a}]\) for the integer $\sem{t}^\I$.
In case that $\phi$ is a sentence and $t$ is a ground term, we shortly
write
$\A\models\phi$ instead of $\A\models\phi[\emptytuple]$, and we write $t^{\A}$
instead of $t^\A[\emptytuple]$.

The \emph{size} \(\abs{\xi}\) of
an \(\FOunC\)-expression \(\xi\) is defined as
the length of the string representation of \(\xi\),
where integers and variables are considered as having length \(1\).
For \(m,q \in \NN\), we write \(\FOunC[\sigma, m, q]\) to denote the set of all
\(\FOunC[\sigma]\)-expressions of size at most \(q\) and with at most \(m\) free variables.

\paragraph*{Hypotheses}
Let $\A$ be a $\sigma$-structure with universe $A$, let $t(\ov{x},\ov{y})$ be an
$\FOunC[\sigma]$-term, let $k\deff |\ov{x}|\geq 1$, let
$\ell\deff|\ov{y}|\geq 0$, and let $\ov{w}\in A^\ell$.
The \emph{hypothesis represented in $\A$ by the counting
term $t(\ov{x},\ov{y})$ and the parameter $\ov{w}$} is defined as the
mapping $f \colon A^k\to \ZZ$ such that $f(\ov{a}) = t^{\A}[\ov{a},\ov{w}]$
for every $\ov{a}\in A^k$.
That is, evaluating the counting term $t$ in
$\A$ while assigning the variables $\ov{x}$ to the elements $\ov{a}$
and assigning the variables $\ov{y}$ to the elements $\ov{w}$
yields the integer $f(\ov{a})$. Henceforth, we will write
$h^{\A}_{t,\ov{w}}$ to denote this function $f$.
For a set \(S \subseteq A^k \times \ZZ\),
we say that \emph{\(h^{\SA}_{t,\ov{w}}\) is consistent with \(S\)}
if and only if \(h^{\SA}_{t, \ov{w}}(\bar{a}) = i\)
for all \((\bar{a}, i) \in S\).

\section{Learning \texorpdfstring{\(\FOunC\)}{FOC1}-Definable Aggregate Queries}
\label{sec:learning}
Let \(\sigma, \sigma'\) be signatures with \(\sigma \subseteq \sigma'\),
fix two numbers
$k\in\NNpos$, $\ell\in\NN$,
and let \(T \subseteq \FOunC[\sigma], T' \subseteq \FOunC[\sigma']\)
be two sets of terms \(t(\bar{x}, \bar{y})\)
with \(\abs{\bar{x}} = k\) and \(\abs{\bar{y}} = \ell\).
We study the following problem.

\begin{framed}
  \noindent
  \(\LearnWithPrecomp\)

  \noindent
  \textbf{Precomputation:}
  Given a \(\sigma\)-structure \(\SA\) with universe \(A\),
  compute a \(\sigma'\)-expansion \(\SA'\) of \(\SA\)
  and a lookup table whose size is independent of $\A$.

  \noindent
  \textbf{Input:}
  Training set \(S \subseteq A^k \times \ZZ\).

  \noindent
  \textbf{Task:}
  Return a term \(t' \in T'\) and a tuple \(\bar{w}' \in A^\ell\)
  such that the hypothesis \(h^{\SA'}_{t', \bar{w}'}\) is consistent with \(S\).
  The algorithm may reject if there is no term \(t \in T\) and tuple \(\bar{w} \in A^\ell\)
  such that the hypothesis \(h^\SA_{t, \bar{w}}\) is consistent
  with \(S\).
\end{framed}

As described in the introduction,
the precomputation phase can be viewed as building an index structure
for the given database $\A$. Afterwards, this index structure can be
used each time that we receive as input a new training set $S$.
The ``learning phase'' is what happens after receiving such a set $S$;
the desired output is a hypothesis that is consistent with $S$.
The main contribution of this paper is an efficient solution for the problem $\LearnWithPrecomp$.
Before presenting the exact statement of our result (\cref{thm:learn-foc1}),
recall from \cref{sec:intro} the discussion on the benefits of
\emph{local-access} algorithms.
In the learning phase (\ie, when receiving a training set),
the algorithm we provide does not require random access to the database.
Instead, it only needs \emph{local access}, i.e., it only accesses the
neighbours of elements that it already holds in memory,
initially starting with the elements given in the training set.
Here, ``neighbours'' refers to neighbours in the Gaifman graph
\(G_\SA\) of the database $\A$.
Formally, \emph{local access} means that the algorithm can access an oracle that answers
requests of the form
``Is $\bar{v} \in R(\SA)$?''
in constant time
and requests of the form
``Return a list of all neighbours of $v$ in $\SA$''
in time linear in the number of neighbours of $v$.
As our machine model,
we use a random-access machine (RAM) model,
and we consider running times under the uniform-cost measure.
This allows us to store an element of the database in a single memory cell
and access it in a single computation step.

The main result of this paper is the following theorem.
\begin{theorem}
  \label{thm:learn-foc1}
  Let $\sigma$ be a signature,
  let $k\in\NNpos$, let $\ell,q\in\NN$,
  let $I$ be a finite set of integers, and
  let $T$ be the set of all  $\FOunC[\sigma, k{+}\ell, q]$-terms that only use integers from \(I\).

  There is an extension $\sigma'$ of $\sigma$ with relation symbols of arity $\leq 1$, and
  there is a number $q'\in\NN$ such that,
  for the set $T'$ of all $\FOunC[\sigma', k{+}\ell, q']$-terms,
  there is an algorithm that solves the problem \(\LearnWithPrecomp\) as follows.

  For a $\sigma$-structure $\SA$ of size $n$ and degree $d$, the
  precomputation to compute the \(\sigma'\)-expansion \(\SA'\) of $\SA$
  and the associated lookup table takes time \(d^{\bigO(1)} {\cdot} n\).
  Afterwards, upon input of a training set $S$ of size $s$, the algorithm
  uses only local access to \(\SA'\),
  access to the lookup table,
  and time \((s + d)^{\bigO(1)}\) to
  return either a hypothesis consistent with $S$ or the answer ``reject''.
  Furthermore, the returned hypothesis can be evaluated
  in time \(d^{\bigO(1)}\),
  using only local access to $\A'$ and access to the lookup table.
\end{theorem}

In particular, this implies that when \(\SA\) comes from a class of
\(\sigma\)-structures of polylogarithmic degree,
then the precomputation takes time quasi-linear in \(n\)
(\ie, $n\cdot (\log n)^{O(1)}$),
a hypothesis is found within time polynomial in \(s\) and polylogarithmic in \(n\),
and it can be evaluated in time polylogarithmic in \(n\).

We remark that the algorithm given in the proof of \cref{thm:learn-foc1}
is not meant to be implemented and used in practice.
Instead, \cref{thm:learn-foc1} serves as a fundamental result
that shows the theoretic (and asymptotic)
feasibility of learning aggregate queries definable in $\FOunC$.
This is in line with previous work on the descriptive complexity of learning
\cite{GroheTuran_Learnability,GroheRitzert_FO,GroheLoedingRitzert_MSO,GrienenbergerRitzert_Trees,vanBergerem_FOCN,vanBergeremSchweikardt_FOWA,vanBergeremGroheRitzert_Parameterized,vanBergerem_PhDThesis}
and closely related work on so-called algorithmic meta theorems
(see, \eg, \cite{Grohe_LogicGraphsAlgorithms,Kreutzer_MetaTheorems2011}).

Before turning to the proof of \cref{thm:learn-foc1},
let us first briefly discuss the situation.
In the assumption of the theorem, we require the set \(I\) of integers
occurring in terms of \(T\) to be finite.
However, note that we can still represent other integers in \(T\)
by using rule~\eqref{def:foc-plustimesterm} for \(\FOunC\),
that is, addition and multiplication.
For example, we could let \(I\) be the set of powers of 10 up to some bound,
and we could obtain the numbers in between with rather few additions and multiplications.
Moreover, requiring \(I\) to be finite does not limit the use
of counting terms of the form \(\FOCCount{\bar{x}}{\phi}\),
which may evaluate to arbitrary integers based on the given database.
Meanwhile, we could let \(I\) contain (large) constants
that are specific to the field the data is coming from,
which allows using them even if the bound \(q\)
on the size of the terms in \(T\) is small.

Since \(I\) is finite and the size of terms in \(T\) is bounded by \(q\),
up to equivalence, the set \(T\) is \emph{finite}.
Thus, when given a $\sigma$-structure $\SA$ with universe $A$
and a training set $S\subseteq A^k\times\ZZ$,
in order to find a hypothesis that is consistent with $S$,
we could proceed as follows.
Loop through all terms $t(\ov{x},\ov{y})$ in $T$.
For each such term $t$, loop through all tuples $\ov{w}\in A^\ell$,
and check whether $t^{\A}[\ov{v},\ov{w}]=\lambda$ for every $(\ov{v},\lambda)\in S$.
If so, stop and output $t$ and $\ov{w}$,
indicating that $h^{\A}_{t,\ov{w}}$ is a hypothesis consistent with $S$.
If no such hypothesis is found, then stop and output ``reject''.
This obviously solves the learning problem.
However, the time taken to loop through all the $\ov{w}\in A^\ell$ is
polynomial in $\abs{A}$,
and it may not suffice to start with the vertices given in the training set
and repeatedly iterate over the neighbours of already found vertices.
Note that \cref{thm:learn-foc1} yields a much more efficient algorithm,
which runs in time polynomial in the size of the training set and the degree of the structure.
This can be substantially faster than being polynomial in the size of the structure.
We achieve this by moving over to a larger signature $\sigma'$ and a suitable $\sigma'$-expansion $\A'$ of $\A$.
This is done in such a way that afterwards,
we only have to loop through those tuples $\ov{w}$ that are close to the tuples $\ov{v}\in A^k$
that actually occur in the training set $S$.
Meanwhile, the number of terms to check is not constant any more, but it depends on $\A$.
However, as we discuss in the proof of \cref{thm:learn-foc1},
this number is polynomial in the degree of $\A$,
which yields the desired running time.
The exact details are provided by the following \cref{lem:equivalent-term};
this lemma is the main technical tool that allows us to prove \cref{thm:learn-foc1}.

For formulating the lemma, we need the following notation.
In a structure \(\SA\)
with universe $A$ and
of degree at most \(d\),
for every \(v \in A\) and any radius \(r \in \NN\),
we have \(\big\lvert\neighbr{\SA}{v}\big\rvert \leq \nu_d(r)
\deff 1 + d \cdot \sum_{0 \leq i < r} (d-1)^i\).
Note that \(\nu_0(r) = 1\),
\(\nu_1(r) \leq 2\),
\(\nu_2(r) \leq 2r+1\),
and \(\nu_d(r) \leq d^{r+1}\)
for all \(r \in \NN\) and \(d \geq 3\).
In particular, for a fixed radius \(r \in \NN\),
\(\nu_d(r)\) is polynomial in \(d\).
For all \(r \in \NN\),
it is straightforward to construct an \(\FO[\sigma]\)-formula
\(\dist_{\leq r}^\sigma (x,y)\) such that for every \(\sigma\)-structure \(\SA\)
and all \(v,w \in A\),
we have \(\SA \models \dist_{\leq r}^\sigma[v,w]\) if and only if
\(\dist^\SA(v,w) \leq r\).
To improve readability, we write
\(\dist^\sigma(x,y) \,{\leq}\, r\) instead of \(\dist_{\leq r}^\sigma (x,y) \),
and \(\dist^\sigma(x,y) \,{>}\, r\) instead of \(\neg\dist_{\leq r}^\sigma (x,y)\).

Let \(r \in \NN\).
An \(\FOunC[\sigma]\)-formula \(\phi(\bar{x})\) with free variables
\(\bar{x} = (x_1, \dots, x_k)\) is \emph{\(r\)-local (around \(\bar{x}\))}
if for every \(\sigma\)-structure \(\SA\) with universe \(A\) and every tuple
\(\bar{v} = (v_1, \dots, v_k) \in A^k\), we have
\(\SA \models \phi[\bar{v}] \iff \Neighbr{\SA}{\bar{v}} \models \phi[\bar{v}]\).
A formula is \emph{local} if it is \(r\)-local for some \(r \in \NN\).
This notion of local formulas is identical with the one in \cite{FlumGrohe_Parameterized}.
It is very similar to the notion of local formulas by Gaifman \cite{Gaifman},
although we use a semantic notion instead of Gaifman's syntactic notion.

For every
\(k \in \NNpos\), every graph \(G\) with vertex set \([k]\),
and every tuple \(\bar{x} = (x_1, \dots, x_k)\) of \(k\) pairwise distinct variables,
we consider the formula
\[\delta^\sigma_{G, r}(\bar{x}) \deff\!\!
  \Land_{\set{i,j} \in E(G)}\!\!\!\!\! \dist^\sigma(x_i, x_j) \leq r
  \ \ \land
  \Land_{\set{i,j} \not\in E(G)}\!\!\!\!\! \dist^\sigma(x_i, x_j) > r.
\]
Note that \(\SA \models \delta^\sigma_{G, 2r+1}[\bar{v}]\) means that the connected components
of the \(r\)-neighbourhood \(\Neighbr{\SA}{\bar{v}}\) correspond to the connected components
of \(G\).
Clearly, the formula \(\delta^\sigma_{G, 2r+1}(\bar{x})\) is \(r\)-local around its free variables \(\bar{x}\).

For two sets \(A, N\) with \(N \subseteq A\), for \(k \in \NN\),
and two tuples \(\bar{w}, \bar{w}' \in A^k\),
we write \(\bar{w} \tupEqInSet{N} \bar{w}'\)
if \(w_i = w'_i\) for all \(i \in [k]\) with \(w_i \in N\).
Note that this notion is not symmetric:
for tuples \(\bar{w}, \bar{w}'\) with \(\bar{w} \tupEqInSet{N} \bar{w}'\),
the tuple \(\bar{w}'\) may still have an entry \(w'_i \in N\) for some \(i \in [k]\)
while \(w_i \not\in N\), so \(w_i \neq w'_i\).

Our main technical ingredient for the proof of \cref{thm:learn-foc1}
is the following lemma.
Note that in the final statement of the lemma, the graphs $H$ are
\emph{connected} and the formulas $\psi$ are local --- both conditions
are absolutely necessary for our proof of
Theorem~\ref{thm:learn-foc1}.

\begin{lemma}
  \label{lem:equivalent-term}
  Let $\sigma$ be a signature,
  let
  $k\in\NNpos$, let $\ell,q\in\NN$,
  let $t(\ov{x},\ov{y})$ be an $\FOunC[\sigma, k{+}\ell, q]$-term, and
  let $I_t$ be the set of integers that occur in $t$.

  There is an extension $\sigma_t$ of $\sigma$ with relation symbols of arity $\leq 1$, and
  there are numbers $q_t,r_t\in\NN$ such that,
  for every $\sigma$-structure $\SA$ of size $n$ and degree $d$, we can
  compute a \(\sigma_t\)-expansion \(\SA_t\) of $\SA$ in time
  \(d^{\bigO(1)} {\cdot} n\) such that the following is true,
  where $A$ denotes the universe of $\SA$.
For all $s\in\NNpos$, for all $\ov{v}_1,\ldots,\ov{v}_s\in A^k$, and for all $\ov{w}\in A^\ell$,
  there is an $\FOunC[\sigma_t, k{+}\ell, q_t]$-term $t'(\ov{x},\ov{y})$ such that for all
  $\ov{w}'\in A^\ell$ with \(\bar{w} \tupEqInSet{N_t} \bar{w}'\)
  for \(N_t \deff \neighb{(2r_t+1)(\ell+q)}{\SA}{\bar{v}_1, \dots, \bar{v}_s}\),
  we have
\(t^\SA[\bar{v}_i, \bar{w}] = (t')^{\SA'}[\bar{v}_i, \bar{w}']\)
  for all \(i \in [s]\).

  Furthermore, $t'$ is a combination via addition and multiplication
  of integers in $I_t$,
  of integers $i$ with
  $-1 \leq i\leq \bigl(\ell \cdot \nu_d\bigl((2r_t{+}1)q\bigr)\bigr)^q$,
  and of counting terms
  of the form
  \(\FOCCount{\bar{z}'}{(\psi \land \delta^{\sigma_t}_{H, 2r_{t}+1})}\)
  where $\psi$ is an $r_t$-local formula in $\FO[\sigma_t]$ and $H$ is a connected graph.
\end{lemma}

We present the proof of \cref{lem:equivalent-term} in \cref{sec:proof-of-main-lemma}.
Intuitively, the lemma says that we can translate a term $t$ into a term $t'$
over an extended signature such that the new term only needs those parameters
that are close to the examples.
By using this lemma, we can prove \cref{thm:learn-foc1} as follows.

\begin{proof}[Proof of \cref{thm:learn-foc1}]
  Let $\sigma,k,\ell,q,I,T$ be chosen according to the assumption of the theorem.

  Note that, up to logical equivalence, there are only finitely many terms in $T$.
  Thus, w.l.o.g.\ we assume that $T$ is finite
  and that all terms in \(T\) only use
  \(x_1, \dots, x_k, y_1, \dots, y_\ell, z_1, \dots, z_q\) as variables.
  For each term $t(\ov{x},\ov{y})$ in $T$,
  we apply \cref{lem:equivalent-term} to obtain an extension
  $\sigma_t\supseteq \sigma$ and numbers $q_t,r_t\in \NN$.
  W.l.o.g.\ we assume that the sets $(\sigma_t\setminus\sigma)_{t\in T}$ are pairwise disjoint.

  Let \(\sigma' \deff \bigcup_{t \in T} \sigma_t\),
  let \(q' \deff \max_{t \in T} q_t\), and
  let $T'$ be the set of all $\FOunC[\sigma', k{+}\ell, q']$-terms
  using only \(x_1, \dots, x_k, y_1, \dots, y_\ell, z_1, \dots, z_{q'}\)
  as variables.
  Furthermore, let \(r' \deff \max_{t \in T} r_t\).

  Upon input of a $\sigma$-structure $\A$ of size $n$ and degree $d$,
  for each $t\in T$ we use \cref{lem:equivalent-term} to compute a
  $\sigma_t$-expansion $\A_t$ of $\A$ in time
  $d^{\bigO(1)} {\cdot} n$.
  We let $\A'$ be the $\sigma'$-expansion of $\A$
  obtained by
  combining all the structures $\A_t$ for $t\in T$.

  In addition, we also compute a lookup table that stores the value
  \(g^{\SA'} \in \ZZ\) for every ground term \(g\) that occurs in a term in \(T'\)
  and is of the form \(\FOCCount{\bar{z}'}{(\psi \land \delta^{\sigma'}_{H, 2r_t+1})}\),
  where \(\psi\) is an \(r_t\)-local formula in \(\FO[\sigma']\) (for some \(t\in T\))
  and \(H\) is a connected graph.

  \begin{claim}
    \label{claim:MainThm:lookup}
    The lookup table can be computed in time \(d^{\bigO(1)} {\cdot} n\).
  \end{claim}
  \begin{claimproof}
    First note that the number of entries in the lookup table is constant
    and does not depend on \(\SA\),
    because the terms in \(T'\) have size at most \(q'\)
    and only use variables from a fixed set.
    Thus, it suffices to show that every single entry of the lookup table
    can be computed in time \(d^{\bigO(1)} {\cdot} n\).

    Let \(g\) be a ground term that occurs in a term in \(T'\)
    and is of the form \(\FOCCount{\bar{z}'}{(\psi \land \delta^{\sigma'}_{H, 2r_t+1})}\),
    where \(\psi\) is an \(r_t\)-local formula in \(\FO[\sigma']\) (for some \(t\in T\))
    and \(H\) is a connected graph.
    Then \(m \deff \abs{\bar{z}'} \leq q'\).
    Let \(\tilde{r} \deff (2r'+1)m\).

    Since \(H\) is a connected graph and \((2r_t+1)m \leq (2r'+1)m = \tilde{r}\),
    we have \(v_2, \dots, v_m \in \neighb{\tilde{r}}{\SA'}{v_1}\)
    for all \(\bar{v} \deff (v_1, \dots, v_m) \in A^m\)
    with \(\SA' \models \delta^\sigma_{H, 2r_t+1}[\bar{v}]\).
    Hence, to compute \(g^{\SA'}\), we can simply initialise the corresponding entry
    in the lookup table with \(0\),
    iterate over all \(v_1 \in A\)
    and all \(v_2, \dots, v_m \in \neighb{\tilde{r}}{\SA'}{v_1}\),
    and increase the entry in the lookup table by \(1\)
    if and only if \(\SA' \models (\psi \land \delta^{\sigma'}_{H, 2r_t+1})[\bar{v}]\) holds.
    For every
    fixed \(v_1 \in A\), we iterate over at most
    \(\nu_d(\tilde{r})^{m-1} \in d^{\bigO(1)}\) tuples
    $(v_2,\ldots,v_m)$.
    Moreover, since \(\psi\) and \(\delta^{\sigma'}_{H, 2r_t+1}\) are
    $r_t$-local,
    for each tuple \(\bar{v}=(v_1,v_2,\ldots,v_m)\),
    it holds that \(\SA' \models (\psi \land \delta^{\sigma'}_{H, 2r_t+1})[\bar{v}]\)
    if and only if
    \(\Neighbrt{\SA'}{\bar{v}} \models (\psi \land \delta^{\sigma'}_{H, 2r_t+1})[\bar{v}]\).
    The latter can be checked by building the neighbourhood structure around \(\bar{v}\)
    in time polynomial in \(d\),
    and then evaluating the formula (of constant size) on the neighbourhood structure
    by a brute-force algorithm in time polynomial in \(d\).

    All in all, we use \(d^{\bigO(1)} {\cdot} n\) iterations,
    and every iteration takes time \(d^{\bigO(1)}\),
    so the overall running time is \(d^{\bigO(1)} {\cdot} n\).
  \end{claimproof}

  Now let us assume that we receive an arbitrary training set
  $S=\set{(\ov{v}_1,\lambda_1),\ldots,(\ov{v}_s,\lambda_s)}\subseteq A^k\times \ZZ$ of size $s$.
Let $T^{*}$ be the set of all those terms $t'(\ov{x},\ov{y})$ in $T'$
  that satisfy the following: $t'$ is a combination via addition and
  multiplication
  of integers in $I$,
  of integers $i$ with
  $-1\leq i\leq \bigl(\ell \cdot \nu_d\bigl((2r'{+}1)q\bigr)\bigr)^q$,
  and of counting terms
  of the form
  \(\FOCCount{\bar{z}'}{(\psi \land \delta^{\sigma'}_{H, 2r_t+1})}\),
  where $\psi$ is an $r_t$-local formula in $\FO[\sigma']$ (for some
  $t\in T$) and $H$ is a connected graph.

  \begin{restatable}{claim}{claimBoundsOnTermEvaluation}\label{claim:MainThm:claim1}
    $T^*$ is of size $d^{\bigO(1)}$.
    Furthermore, for every $t'(\ov{x},\ov{y}) \in T^*$, upon input of
    $\ov{a}\in A^k$ and $\ov{b}\in A^\ell$, we can compute
    $(t')^{\A'}[\ov{a},\ov{b}]$ in time $d^{\bigO(1)}$
    with only local access to \(\SA'\)
    and access to the precomputed lookup table.
  \end{restatable}

  \begin{claimproof}
    Since the terms in \(T^*\) have length at most \(q'\),
    the variables come from a fixed finite set,
    the signature \(\sigma'\) has constant size,
    and the integers that may occur in a term in \(T^*\)
    come from a set of size polynomial in \(d\),
    the total number of terms in \(T^*\) is also
    polynomial in \(d\).

    For the evaluation of a term $t'$ in $T^*$,
    we first consider counting terms \(t'(\bar{x}', \bar{y}')\) of the form
    \(\FOCCount{\bar{z}'}{(\psi \land \delta^{\sigma'}_{H, 2r_t+1})}\),
    where
    \(k' \deff \abs{\bar{x}'} \leq k\),
    \(\ell' \deff \abs{\bar{y}'} \leq \ell\),
    \(m' \deff \abs{\bar{z}'} \leq q'\),
    $\psi$ is an $r_t$-local formula in $\FO[\sigma']$ (for some \(t \in T\)),
    and $H$ is a connected graph with
    \(V(H) = [k' + \ell' + m']\).

    In case that  $k'{+}\ell' = 0$, the term $t'$ is a ground term. Hence,
    we can simply use the precomputed lookup table
    to get the value $(t')^{\A'}\in\ZZ$ in time $O(1)$.

    In case that $k'{+}\ell'>0$,
    let \(\bar{a} \in A^{k'}\) and \(\bar{b} \in A^{\ell'}\).
    Since \(H\) is connected and \(r_t \leq r'\),
    for all \(\bar{c} = (c_1, \dots, c_{m'})\) with
    \(\SA' \models \delta^{\sigma'}_{H,2r_t+1}[\bar{a}, \bar{b}, \bar{c}]\),
    we have
    \(c_1, \dots, c_{m'} \in \neighb{(2r'+1)m'}{\SA'}{\bar{a},\bar{b}}\).

    Thus, for $\tilde{r}\deff (2r'{+}1)m'\leq (2r'{+}1)q'$, we have
    \begin{align*}
      (t')^{\SA'}[\bar{a}, \bar{b}]
      &= \abs{\bigsetc{\bar{c} \in A^{m'}}
      {\SA' \models (\psi \land \delta^{\sigma'}_{H,2r_t+1})[\bar{a},
        \bar{b}, \bar{c}]}}\\
      &= \abs{\bigsetc{\bar{c} \in \bigl(\neighb{\tilde{r}}{\SA'}{\bar{a},\bar{b}}\bigr)^{m'}}
      {\SA' \models (\psi \land \delta^{\sigma'}_{H,2r_t+1})[\bar{a}, \bar{b}, \bar{c}]}}.
    \end{align*}
    Furthermore, \(\psi\) and \(\delta^{\sigma'}_{H,2r_t+1}\) are
    $r_t$-local formulas,
    so
    \(\SA' \models (\psi \land \delta^{\sigma'}_{2r_t+1,H})[\bar{a}, \bar{b}, \bar{c}]\)
    if and only if
    \(\Neighb{r_t}{\SA'}{\bar{a},\bar{b},\bar{c}} \models (\psi \land \delta^\sigma_{2r_t+1,H})[\bar{a}, \bar{b}, \bar{c}]\).
    Since the size of the neighbourhood is polynomial in \(d\),
    the evaluation of \((t')^{\SA'}[\bar{a}, \bar{b}]\)
    can be performed by evaluating an \(\FO[\sigma']\)-formula of constant size
    on a structure of size polynomial in \(d\)
    for a number of assignments that is polynomial in \(d\).
    The evaluation of the formula can be done by building the
    neighbourhood structure around $\bar{a},\bar{b}$ in time polynomial in \(d\),
    and then evaluating the formula on the neighbourhood structure
    by a brute-force algorithm in time polynomial in \(d\).

    Now let \(t'\) be an arbitrary term in \(T^*\).
    Then, for all \(\bar{a} \in A^k\), \(\bar{b} \in A^\ell\),
    \((t')^{\SA'}[\bar{a}, \bar{b}]\) can be evaluated in time polynomial in \(d\)
    with only local access to \(\SA'\)
    and access to the precomputed lookup table
    by first evaluating every counting term in \(t'\) as described above
    and then combining the results and the integers occurring in \(t'\)
    via a constant number of additions and multiplications.
  \end{claimproof}

  To find a hypothesis that is consistent with $S$, we loop through all
  terms $t'(\ov{x},\ov{y})$ in $T^*$. For each such term $t'$, we loop
  through all tuples $\ov{w}'\in N^\ell$ for
  \(N \deff \neighb{(2r'+1)(\ell+q)}{\SA}{\bar{v}_1, \dots, \bar{v}_s}\).
  We check if $(t')^{\A'}[\ov{v}_i,\ov{w}']=\lambda_i$ for all
  $i\in[s]$.
  If so, we stop and output $t'$ and $\ov{w}'$, indicating that
  $h^{\A'}_{t',\ov{w}'}$ is a hypothesis consistent with $S$.
  Otherwise, we stop and output ``reject''.

  \begin{claim}\label{claim:MainThm:claim2}
    This algorithm uses only local access to \(\SA'\)
    and access to the precomputed lookup table, and
    it terminates within time $(s+d)^{\bigO(1)}$.
    If it outputs a hypothesis, this hypothesis is consistent with $S$.
    If it outputs ``reject'', there is no term $t\in T$ and tuple
    $\ov{w}\in A^\ell$ such that $h^{\A}_{t,\ov{w}}$ is consistent with $S$.
  \end{claim}
  \begin{claimproof}
    The set \(N\) contains at most \(s \cdot k \cdot \nu_d\bigl((2r'{+}1)(\ell{+}q)\bigr)\)
    elements, which is polynomial in \(s\) and \(d\).
    By \cref{claim:MainThm:claim1},
    \(T^*\) is of size \(d^{\bigO(1)}\).
    Thus, in total, we iterate over \((s+d)^{\bigO(1)}\) hypotheses.
    For every hypothesis consisting of a term \(t' \in T^*\) and a tuple \(\bar{w}' \in A^\ell\),
    by \cref{claim:MainThm:claim1}, we can check in time \(d^{\bigO(1)} {\cdot} s\)
    if \((t')^{\SA'}[\bar{v}_i, \bar{w}'] = \lambda_i\) for all
    \(i\in[s]\).
    This check uses only
    local access to \(\SA'\) and access to the precomputed lookup
    table.
    This proves the first statement of the claim.
The second statement of the claim is obvious.

    To prove the third statement of the claim, let us assume that there exists
    a term $t\in T$ and a tuple $\ov{w}\in A^\ell$ such that
    $h^{\A}_{t,\ov{w}}$ is consistent with $S$.
    For this particular choice of $t$ and $\ov{w}$, and for the given
    tuples $\ov{v}_1,\ldots,\ov{v}_s$,
    \cref{lem:equivalent-term}
    yields an $\FOunC[\sigma_t, k{+}\ell, q_t]$-term $t'(\ov{x},\ov{y})$
    such that
    for all
    $\ov{w}'\in A^\ell$ with \(\bar{w} \tupEqInSet{N_t} \bar{w}'\)
    for
    \(N_t \deff \neighb{(2r_t+1)(\ell+q)}{\SA}{\bar{v}_1, \dots, \bar{v}_s}\),
    we have
    \[
      t^\SA[\bar{v}_i, \bar{w}] \ = \ (t')^{\SA'}[\bar{v}_i, \bar{w}']\quad \text{for all \(i \in [s]\)}.
    \]
    Note that $N_t\subseteq N$. Hence, our algorithm will consider at least one
    $\ov{w}'\in N^\ell$ such that $\ov{w} \tupEqInSet{N_t}
    \ov{w}'$. Let us fix such a $\ov{w}'$.
    All that remains to be done is to show that $t'\in T^*$ --- this will
    imply that our algorithm
    will
    eventually consider $t',\ov{w}'$
    and then stop and output $t'$ and $\ov{w}'$, indicating that
    $h^{\A'}_{t',\ov{w}'}$ is a hypothesis consistent with $S$.

    By our choice of $\sigma'$ and $q'$, we know that $\sigma'\supseteq
    \sigma_t$ and $q'\geq q_t$. Thus, $t'\in T'$.
    From \cref{lem:equivalent-term}, we
    know that
    $t'$ is a combination via addition and multiplication
    of integers in $I$,
    of integers $i$ with
    $-1 \leq i\leq \bigl(\ell \cdot \nu_d\bigl((2r_t{+}1)q\bigr)\bigr)^q$, and
    of counting terms
    of the form \(\FOCCount{\bar{z}'}{(\psi \land \delta^{\sigma'}_{H, 2r_{t}+1})}\),
    where $\psi$ is an $r_t$-local formula in $\FO[\sigma_t]$ and $H$ is a
    connected graph.
    By our particular choice of $T^*$ and since $r'\geq r_t$, we obtain
    that $t'\in T^*$.
    This completes the proof of Claim~\ref{claim:MainThm:claim2}.
  \end{claimproof}

  In summary, the proof of \cref{thm:learn-foc1} is complete.
\end{proof}
\section{Proof of Lemma~\ref{lem:equivalent-term}}
\label{sec:proof-of-main-lemma}
The proof of \cref{lem:equivalent-term} heavily relies on the following \emph{localisation theorem
for \(\FOunC\)}.
This theorem is implicit in
\cite{GroheSchweikardt_FOunC}; here we present it in a way analogous
to \cite[Theorem~4.7]{vanBergeremSchweikardt_FOWA}.

\begin{theorem}[Localisation Theorem for \(\FOunC\), \cite{GroheSchweikardt_FOunC}]
  \label{thm:foc1-localisation}
  Let \(k \in \NN\), and let \(\sigma\) be a signature.
  For every \(\FOunC[\sigma]\)-formula \(\phi(x_1, \dots, x_k)\),
  there is an extension \(\sigma_\phi\) of \(\sigma\)
  with relation symbols of arity \(\leq 1\),
  and an \(\FO[\sigma_\phi]\)-formula
  \(\phi'(x_1, \dots, x_k)\) that is a Boolean combination of local formulas
  and statements of the form \(R()\) where \(R \in \sigma_\phi\) has arity \(0\),
  for which the following holds.
There is an algorithm that,
  upon input of a \(\sigma\)-structure \(\SA\) of size $n$ and degree
  $d$,
  computes in time
  \(d^{\bigO(1)} {\cdot} n\)
  a \(\sigma_\phi\)-expansion \(\SA_{\phi}\) of \(\SA\)
  such that for all
  \(\bar{v} \in A^k\) (where $A$ denotes the universe of $\SA$),
  we have \(\SA_\phi \models \phi'[\bar{v}]\)
  $\iff$
  \(\SA \models \phi[\bar{v}]\).
\end{theorem}

For the proof of \cref{lem:equivalent-term},
let $\sigma,k,\ell,q,t(\ov{x},\ov{y}),I_t$ be chosen according to the
assumption of the lemma. In particular, $t(\ov{x},\ov{y})$ is an
\(\FOunC[\sigma, k{+}\ell, q]\)-term, and $I_t$ is the set of integers
that occur in $t$.

We first note that
it suffices to prove the statement of the lemma for
the particular case where \(t(\bar{x}, \bar{y})\)
is of the form \(\FOCCount{\bar{z}}{\phi(\bar{x}, \bar{y}, \bar{z})}\)
for some \(\FOunC[\sigma]\)-formula \(\phi\).
Assume for now that the statement of the lemma holds for all terms of this
form, and consider an arbitrary \(\FOunC[\sigma, k{+}\ell, q]\)-term $t(\ov{x},\ov{y})$ that is
\emph{not} of this form.
Then, \(t\) is a combination via addition and multiplication of integers from \(I_t\)
and of counting terms \(u\) of the form \(\FOCCount{\bar{z}}{\phi(\bar{x}, \bar{y}, \bar{z})}\)
for some \(\FOunC[\sigma]\)-formula \(\phi\).
Let \(U\) be the set of all these counting terms \(u\) occurring in
$t$.
We assume that the statement of the lemma holds for each $u\in U$.
Hence, for each $u\in U$, we obtain
an extension \(\sigma_u\) of \(\sigma\) with relation symbols of arity \(\leq 1\)
and numbers \(q_u, r_u \in \NN\).
W.l.o.g.~we assume that the sets \((\sigma_u \setminus \sigma)_{u \in U}\)
are pairwise disjoint.
Let \(\sigma_t \deff \bigcup_{u \in U} \sigma_u\),
\(r_t \deff \max_{u \in U} r_u\), and
\(q_t \deff q\cdot \max_{u \in U} q_u\).

Upon input of a \(\sigma\)-structure \(\SA\) of size \(n\) and degree \(d\),
for each  \(u \in U\),
the statement of the lemma for $u$ enables us to
compute a \(\sigma_u\)-expansion \(\SA_u\) of \(\SA\)
in time \(d^{\bigO(1)} {\cdot} n\).
We let \(\SA_t\) be the \(\sigma_t\)-expansion of \(\SA\)
obtained by combining all the structures \(\SA_u\) for all counting
terms \(u \in U\).
Let $A$ denote the universe of $\A$.

When given $\ov{v}_1,\ldots,\ov{v}_s\in A^k$ and $\ov{w}\in A^\ell$,
the statement of the lemma for $u$ yields
an \(\FOunC[\sigma_u, k{+}\ell, q_u]\)-term \(u'\) for each $u\in U$.
We choose \(t'\) to be the term obtained from \(t\)
by replacing every occurrence of a counting term \(u \in U\)
by the corresponding counting term \(u'\).
Clearly $t'$ has length at most $q_t$.
It is not difficult to verify that $t'$ has the desired properties
stated in \cref{lem:equivalent-term}.

All that remains to be done is to prove the statement of the lemma for the
particular case where
\(t(\bar{x}, \bar{y})\)
is of the form \(\FOCCount{\bar{z}}{\phi(\bar{x}, \bar{y}, \bar{z})}\)
for some \(\FOunC[\sigma]\)-formula \(\phi\).
Given such a term $t$, let \(m \deff \abs{\bar{z}}\).
Using \cref{thm:foc1-localisation},
we obtain an extension \(\sigma'\deff\sigma_\phi\) of \(\sigma\) with relation symbols of arity \(\leq 1\)
and an \(\FO[\sigma']\)-formula \(\phi'(\bar{x}, \bar{y}, \bar{z})\)
that is a Boolean combination of local formulas and statements of the form \(R()\),
where \(R \in \sigma'\) has arity \(0\),
for which the following holds.
Given a \(\sigma\)-structure \(\SA\) of size \(n\) and degree \(d\)
and with universe \(A\),
we can compute a \(\sigma'\)-expansion \(\SA'\deff\SA_\phi\) of \(\SA\) in time
\(d^{\bigO(1)} {\cdot} n\) such that
for all \(\bar{a} \in A^k\), \(\bar{b} \in A^\ell\), and \(\bar{c} \in A^m\), we have
$\SA\models\phi[\bar{a},\bar{b},\bar{c}]$ $\iff$
$\SA'\models \phi'[\bar{a},\bar{b},\bar{c}]$.
Thus, for
\(\tilde{t}(\ov{x},\ov{y}) \deff
\FOCCount{\bar{z}}{\phi'(\bar{x}, \bar{y}, \bar{z})}\),
we have $\tilde{t}^{\A'}[\ov{a},\ov{b}]=t^\A[\ov{a},\ov{b}]$ for
all $\ov{a}\in A^k$, $\ov{b}\in A^\ell$.

From the particular shape of $\phi'$, we obtain that there
exists a number $r' \in\NN$ such that $\phi'$ is $r'$-local, \ie,
for all \(\bar{a} \in A^k\), \(\bar{b} \in A^\ell\), \(\bar{c} \in A^m\),
we have
$\SA'\models \phi'[\bar{a},\bar{b},\bar{c}]$ $\iff$
$\NeighbrStrich{\SA'}{\bar{a},\bar{b},\bar{c}}\models \phi'[\bar{a},\bar{b},\bar{c}]$.

Let $\CG$ be the set of undirected graphs with vertex set \([k + \ell + m]\).
For every $G\in \CG$, consider the formula
\(\phi'_G(\bar{x}, \bar{y}, \bar{z}) \deff \phi'(\bar{x}, \bar{y}, \bar{z})
\ \land \ \delta^{\sigma'}_{G, 2 r' + 1} (\bar{x}, \bar{y},\bar{z})\)
and the term
\(u_G(\bar{x}, \bar{y}) \deff \FOCCount{\bar{z}}{\phi'_G(\bar{x}, \bar{y}, \bar{z})}\).
It is not difficult to verify that for all
\(\bar{a} \in A^k\), \(\bar{b} \in A^\ell\), we have
\[
  t^\SA[\bar{a}, \bar{b}]
  \ \ = \ \
  \tilde{t}^{\SA'}[\bar{a},\bar{b}]
  \ \ = \ \
  \sum_{G \in \CG} \
  (u_G)^{\SA'}[\bar{a}, \bar{b}].
\]

To further decompose the terms \(u_G\),
we use techniques similar to the ones used in \cite{GroheSchweikardt_FOunC}
to decompose terms into so-called connected local terms.
With these, we obtain the following technical lemma.
The statement of this lemma as well as its proof are highly
non-trivial. It depends on a careful analysis of the connected
components of undirected graphs with $k{+}\ell{+}m$ nodes.
Note that in the final statement of the lemma, the graphs $H$ are
\emph{connected} and the formulas $\psi$ are local --- both conditions
are also stated in \cref{lem:equivalent-term}
and are absolutely necessary for our proof of \cref{thm:learn-foc1}.

\begin{lemma}
  \label{lem:equivalent-term-G}
  Let $\sigma'$ be a signature and
  let $r',k,\ell,m\in\NN$
  with $k{+}\ell{+}m\geq 1$.
  Let $\phi'(\ov{x},\ov{y},\ov{z})$ be an $r'$-local $\FO[\sigma']$-formula
  with $|\ov{x}|=k$, $|\ov{y}|=\ell$, and $|\ov{z}|=m$.
  Let $\CG$ be the set of all undirected graphs with vertex set
  $[k+\ell+m]$. Let $G\in\CG$ and let
  \[
    u_G(\ov{x},\ov{y}) \ \deff \ \
    \FOCCount{\bar{z}}{\bigl(\phi'(\bar{x}, \bar{y}, \bar{z})
    \land \delta^{\sigma'}_{G, 2r'+1}(\bar{x}, \bar{y}, \bar{z})\bigr)}.
  \]
  There is a number $q_G\in\NN$ such that,
  for every $\sigma'$-structure $\A'$ of degree $d$ and with universe $A$,
  for all $s\in\NNpos$, for all $\ov{v}_1,\ldots,\ov{v}_s\in A^k$, and
  for all $\ov{w}\in A^\ell$, there exists an
  \(\FOunC[\sigma', k{+}\ell, q_{G}]\)-term \(u'_G(\bar{x}, \bar{y})\)
  such that the following is true for
  \(N \deff \neighb{(2r'+1)(\ell+m)}{\SA'}{\bar{v}_1, \dots, \bar{v}_s}\).

  For all \(\bar{w}' \in A^\ell\) with \(\bar{w} \tupEqInSet{N} \bar{w}'\),
  we have
  \[
    (u_G)^{\SA'}[\bar{v}_i, \bar{w}]
    \ \, = \ \,
    (u'_G)^{\SA'}[\bar{v}_i, \bar{w}']
  \quad\text{for all \(i \in [s]\).}
  \]
  Furthermore, \(u'_G\) is a combination via addition and multiplication of integers \(i\)
  with \(-1 \leq i \leq \Bigl(\ell \cdot \nu_d\bigl((2r'{+}1)m\bigr)\Bigr)^m\)
  and of counting terms of the form
  \(\FOCCount{\bar{z}'}{(\psi \land \delta^{\sigma'}_{H, 2r'+1})}\),
  where \(\abs{\bar{z}'} \leq \abs{\bar{z}}\),
  for a connected graph \(H\) and an \(r'\)-local formula \(\psi\).
\end{lemma}

Before proving Lemma~\ref{lem:equivalent-term-G}, we first finish the proof of \cref{lem:equivalent-term}.
We apply \cref{lem:equivalent-term-G} to the term \(u_G\) for all
$G\in\CG$. For each $G\in\CG$ this yields a number $q_G\in\NN$.
We choose $r_t\deff r'$ and $\sigma_t\deff\sigma'$.

For any $\sigma$-structure $\A$, we let $\A'$ be the
$\sigma'$-expansion obtained as described above (by applying
Theorem~\ref{thm:foc1-localisation} to the formula $\phi$).
When given tuples $\ov{v}_1,\ldots,\ov{v}_s\in A^k$ and $\ov{w}\in
A^\ell$, we obtain from \cref{lem:equivalent-term} an
$\FOunC[\sigma', k{+}\ell, q_{G}]$-term \(u'_G(\bar{x}, \bar{y})\),
for every $G\in\CG$.
This term \(u'_G\) is a combination via addition and multiplication of integers \(i\)
with \(-1 \leq i \leq \Bigl(\ell \cdot \nu_d\bigl((2r'{+}1)m\bigr)\Bigr)^m\)
and of counting terms of the form \(\FOCCount{\bar{z}'}{(\psi \land \delta^{\sigma'}_{H, 2r'+1})}\),
where \(\abs{\bar{z}'} \leq \abs{\bar{z}}\),
for a connected graph \(H\) and an \(r'\)-local formula \(\psi\).
We choose
\[
  t'(\ov{x},\ov{y}) \ \deff \ \ \sum_{G\in\CG} \ u'_G(\ov{x},\ov{y})
\]
and let $q_t\deff \sum_{G\in\CG} (q_G+3)$.
Note that $t'$ has length $\leq q_t$ and that \(m \leq q\).
It is not difficult to verify that the term $t'$ has the desired
properties stated in \cref{lem:equivalent-term}.

All that remains to be done to complete the proof of
\cref{lem:equivalent-term} is
to prove \cref{lem:equivalent-term-G}.

\begin{proof}[Proof of \cref{lem:equivalent-term-G}]
  We prove the statement by induction on the number \(c\) of connected components of \(G\).

  Consider the induction base \(c = 1\), \ie, the graph \(G\) is connected.
  We choose $q_G$ to be the length of the term $u_G$.
  For choosing the term $u'_G$, we distinguish between 4~cases.
\begin{description}
    \item[Case 1:]
      \(k = \ell = 0\).
      In this case, we set \(u'_G \deff u_G\).
    \item[Case 2:]
      \(k = 0\) and \(\ell >0\).
      In this case, we set \(u'_G \deff i\) for the particular number
      $i\deff (u_G)^{\SA'}[\bar{w}] \in \ZZ$.
      Since \(G\) is connected, all elements in a tuple \(\bar{c} \in A^m\)
      with \(\SA' \models \delta^{\sigma'}_{G, 2r'+1}[\bar{w}, \bar{c}]\)
      have distance at most \((2r'{+}1) m\) from \(\bar{w}\).
      Therefore, \(0 \leq i \leq \bigl(\ell \cdot
      \nu_d\bigl((2r'{+}1)m\bigr)\bigr)^m\).
    \item[Case 3:]
      \(k >0 \) and \(\bar{w} \not\in N^\ell\).
      Since $G$ is connected, for all $i\in[s]$ and all \(\bar{c} \in A^m\),
      it holds that
      \(
        \SA' \ \not\models \ \delta^{\sigma'}_{G, 2r'+1}[\bar{v}_i, \bar{w},
        \bar{c}].
      \)
      Hence, for all \(i \in [s]\) we have
      \((u_G)^{\SA'}[\bar{v}_i, \bar{w}] = 0\).
      Therefore, we can choose \(u'_G \deff 0\).
    \item[Case 4:]
      \(k >0\) and \(\bar{w} \in N^\ell\).
      In this case, $\bar{w}$ is the only tuple $\bar{w}'$ that satisfies
      \(\bar{w} \tupEqInSet{N} \bar{w}'\).
      Hence, we can choose \(u'_G \deff u_G\).
  \end{description}
This completes the induction base.

  We now turn to the induction step, where \(c > 1\).
  We assume that the statement of the lemma holds for all graphs \(G'\)
  with fewer than \(c\) connected components.
  Let \(C_1\) be the set of all vertices of \(G\) that are in the same connected component
  as the vertex \(1\).
  Let \(C_2 \deff V(G) \setminus C_1\).
  For each $j\in\set{1,2}$, let $G[C_j]$ be the induced subgraph of $G$ with
  vertex set $C_j$. Clearly, $G$ is the disjoint union of the graphs
  $G[C_1]$ and $G[C_2]$, and
  \(G[C_1]\) has only one connected component,
  and \(G[C_2]\) has \(c{-}1\) connected components.
  For \(j \in \set{1,2}\),
  let \(\bar{x}_j\) be the tuple of variables of \(\bar{x}\)
  such that their index is contained in \(C_j\),
  let \(\bar{y}_j\) be the tuple of variables of \(\bar{y}\)
  such that their index \(+ k\) is contained in \(C_j\),
  and let \(\bar{z}_j\) be the tuple of variables of \(\bar{z}\)
  such that their index \(+ k {+} \ell\) is contained in \(C_j\).

  By using the Feferman-Vaught Theorem
  (cf., \cite{FefermanVaught_1959,Makowsky_FefermanVaught2004,vanBergeremSchweikardt_FOWA}),
  we obtain a decomposition \(\Delta\) of \(\phi'(\bar{x}, \bar{y}, \bar{z})\)
  w.r.t.~\((\bar{x}_1 \bar{y}_1 \bar{z}_1 ; \bar{x}_2 \bar{y}_2
  \bar{z}_2)\).
  \Ie, \(\Delta\) is a finite non-empty set of pairs of \(r'\)-local $\FO[\sigma']$-formulas
  of the form
  \((\alpha(\bar{x}_1, \bar{y}_1, \bar{z}_1), \beta(\bar{x}_2, \bar{y}_2, \bar{z}_2))\)
  such that for all \(\bar{a} \in A^k\),
  \(\bar{b} \in A^\ell\),
  \(\bar{c} \in A^m\), we have
  \(\SA' \models \phi'[\bar{a}, \bar{b}, \bar{c}]\)
  $\iff$ there exists a pair \((\alpha, \beta) \in \Delta\)
  such that \(\SA' \models \alpha[\bar{a}_1, \bar{b}_1, \bar{c}_1]\)
  and \(\SA' \models \beta[\bar{a}_2, \bar{b}_2, \bar{c}_2]\).
  Here, \(\bar{a}_j, \bar{b}_j, \bar{c}_j\) are defined analogously to
  \(\bar{x}_j, \bar{y}_j, \bar{z}_j\) for \(j \in \set{1,2}\).
  Moreover, the pairs in \(\Delta\) are mutually exclusive,
  \ie, for all \(\bar{a} \in A^k\), \(\bar{b} \in A^\ell\), \(\bar{c} \in A^m\),
  there is at most one pair \((\alpha, \beta) \in \Delta\) such that
  \(\SA'\models \alpha[\bar{a}_1, \bar{b}_1, \bar{c}_1]\)
  and \(\SA' \models \beta[\bar{a}_2, \bar{b}_2, \bar{c}_2]\).

  Let \(\CG_{\neg G}\) be the set of graphs \(H \neq G\)
  with vertex set \(V(H) = [k + \ell + m]\)
  and \(G[C_1] = H[C_1]\) and \(G[C_2] = H[C_2]\).
  Note that such graphs \(H\) have strictly fewer connected components than \(G\).

  For every pair \((\alpha, \beta) \in \Delta\), we set
  \begin{align*}
    t_{\alpha, \beta}(\bar{x}, \bar{y})
    &\deff \FOCCount{\bar{z}}{\bigl(\alpha(\bar{x}_1, \bar{y}_1, \bar{z}_1) \land
    \beta(\bar{x}_2, \bar{y}_2, \bar{z}_2) \land
      \delta^{\sigma'}_{G, 2r'+1}(\bar{x}, \bar{y}, \bar{z})\bigr)},
    \smallskip\\
    t_{\alpha, 1}(\bar{x}_1, \bar{y}_1)
    &\deff \FOCCount{\bar{z}_1}{\bigl(\alpha(\bar{x}_1, \bar{y}_1, \bar{z}_1) \land
    \delta^{\sigma'}_{G[C_1], 2r'+1}(\bar{x}_1, \bar{y}_1,
      \bar{z}_1))\bigr)},
    \smallskip\\
    t_{\beta, 2}(\bar{x}_2, \bar{y}_2)
    &\deff \FOCCount{\bar{z}_2}{\bigl(\beta(\bar{x}_2, \bar{y}_2, \bar{z}_2) \land
    \delta^{\sigma'}_{G[C_2], 2r'+1}(\bar{x}_2, \bar{y}_2,
      \bar{z}_2))\bigr)},
    \smallskip\\
    t_{\alpha, \beta, \neg G}(\bar{x}, \bar{y})
    &\deff \sum_{H \in \CG_{\neg G}} \# \bar{z}.
    \bigl(\alpha(\bar{x}_1, \bar{y}_1, \bar{z}_1) \land
    \beta(\bar{x}_2, \bar{y}_2, \bar{z}_2)
    \land \ \delta^{\sigma'}_{H, 2r'+1}(\bar{x}, \bar{y}, \bar{z})\bigr).
  \end{align*}

  Since the pairs \((\alpha, \beta) \in \Delta\) are mutually exclusive,
  for all \(\bar{a} \in A^k\), \(\bar{b} \in A^\ell\), we have
  \(
    (u_G)^{\SA'}[\bar{a}, \bar{b}]
     \ = \
  \sum_{(\alpha, \beta) \in \Delta} t^{\SA'}_{\alpha,
    \beta}[\bar{a}, \bar{b}].
  \)
  Furthermore, it is not difficult to verify that
  \[
    t^{\SA'}_{\alpha, \beta}[\bar{a}, \bar{b}]
    \ = \ \
    t^{\SA'}_{\alpha, 1}[\bar{a}_1, \bar{b}_1] \cdot t^{\SA'}_{\beta, 2}[\bar{a}_2, \bar{b}_2]
     \ \ - \ \ t^{\SA'}_{\alpha, \beta, \neg G}[\bar{a}, \bar{b}],
  \]
  where \(\bar{a}_j, \bar{b}_j\) are defined analogously to \(\bar{x}_j, \bar{y}_j\)
  for \(j \in \set{1,2}\).

  Using the induction hypothesis,
  we apply the statement of the lemma to
  \(t_{\alpha, 1}\),
  \(t_{\beta, 2}\),
  and to every summand of \(t_{\alpha, \beta, \neg G}\),
  and we call the resulting terms
  \(t'_{\alpha, 1}\),
  \(t'_{\beta, 2}\),
  and \(t'_{\alpha, \beta, \neg G}\).

  We set \(t'_{\alpha, \beta}(\bar{x}, \bar{y}) \deff
  t'_{\alpha, 1}(\bar{x}_1, \bar{y}_1) \cdot t'_{\beta, 2}(\bar{x}_2, \bar{y}_2)
  - t'_{\alpha, \beta, \neg G}(\bar{x}, \bar{y})\)
  and choose \(u'_G(\ov{x},\ov{y}) \deff \sum_{(\alpha, \beta) \in
    \Delta} t'_{\alpha, \beta}(\ov{x},\ov{y})\).

  It is not difficult to verify that the term \(u'_G\) has the desired properties.
  Moreover, the size of \(u'_G\) can be bounded by a number \(q_{G}\)
  that only depends on \(u_G\) (but not on $\A'$).
  This completes the proof of \cref{lem:equivalent-term-G}. Hence,
  also the proof of \cref{lem:equivalent-term} now is complete.
\end{proof}

\section{Conclusion}
\label{sec:conclusion}
We have studied the complexity of learning aggregate queries from examples.
For this, we have extended the framework for Boolean classification problems
by Grohe and Tur{\'{a}}n~\cite{GroheTuran_Learnability}
to multiclass classification problems with integer-valued classifiers.
In our setting, such classifiers are represented by a pair $(t,\ov{w})$
where $t(\ov{x},\ov{y})$ is a counting term in $\FOunC$
and $\ov{w}$ is a tuple of elements in the universe \(A\)
of the given database $\A$.

Our main result shows that we can build a suitable index structure on
the given database $\A$ during a
precomputation step whose running time is linear in the size
and polynomial in the degree of $\A$. Afterwards, by utilising this
index structure, whenever receiving
as input a new training set $S$, a
classifier definable in \(\FOunC\)
can be learned in time polynomial in the degree of $\A$
and polynomial in the number of examples given in the training set.
The classifiers returned by our algorithm
can be evaluated efficiently, in time polynomial in the degree of $\A$.
Moreover, after having built the index structure, all our algorithms
require only local access to the database.

It seems conceivable that our results can be generalised to the more expressive logic
\(\FOWAun\) that operates on weighted structures \cite{vanBergeremSchweikardt_FOWA},
since the locality results obtained for this logic in \cite{vanBergeremSchweikardt_FOWA}
are similar to the ones obtained for \(\FOunC\) in
\cite{GroheSchweikardt_FOunC} and used here.

For the \emph{Boolean} classification problem based on $\FO$,
\cite{vanBergeremGroheRitzert_Parameterized}
shows that the learning problem is fixed-parameter tractable on nowhere dense classes.
This result heavily relies
on the fixed-parameter
tractability of the evaluation problem for $\FO$
on nowhere dense classes \cite{GroheKreutzerSiebertz_NowhereDense}.
From \cite{GroheSchweikardt_FOunC} it is known that on nowhere dense
classes also the evaluation problem for formulas and terms of $\FOunC$ is
fixed-parameter tractable.
We therefore think that $\FOunC$ is a good candidate for a logic with
a fixed-parameter tractable \emph{integer-valued} classification problem
on classes of sparse structures.

In this paper, the task in the learning problem is to find an
integer-valued hypothesis
that is consistent with the training set.
Other publications on the considered framework for \emph{Boolean} classification problems
also study settings in which one wants to find a hypothesis that
\emph{generalises well}
\cite{GroheRitzert_FO,GroheLoedingRitzert_MSO,GrienenbergerRitzert_Trees,vanBergerem_FOCN,vanBergeremSchweikardt_FOWA,vanBergeremGroheRitzert_Parameterized,vanBergerem_PhDThesis}.
More specifically, they study \emph{probably approximately correct (PAC)} learning tasks
in which the training examples are considered as being generated from
a (fixed, but unknown) probability distribution.
The task is to find a hypothesis that has a small expected error on new examples
generated from the same distribution.
While PAC-learning algorithms for Boolean classification problems
are typically based on the
empirical risk minimisation (ERM) principle
\cite{Shalev-ShwartzBen-David_UnderstandingMachineLearning},
results in algorithmic learning theory for multiclass classification
\cite{DBLP:conf/focs/BrukhimCDMY22,DBLP:conf/nips/DanielySS12,DBLP:conf/colt/HannekeMZ23}
are based on different principles and even show that
the ERM principle in general does not suffice for PAC learning with an unbounded number of classes
\cite{DanielyShalevShwartz_Multiclass2014}.
Therefore, we expect it to be quite challenging to obtain PAC-learning results
for our framework of multiclass classification problems.

We plan to work on the raised questions in future work.

\bibliography{main}

\end{document}